\documentclass[prl,twocolumn]{revtex4}
\usepackage[titletoc,toc,title]{appendix}
\usepackage[ansinew]{inputenc}
\usepackage{graphicx}
\usepackage{amsmath}
\usepackage{amsthm}
\usepackage{bm}
\usepackage{layout}
\usepackage{float}
\usepackage{amsfonts}
\usepackage{amssymb}
\usepackage{txfonts}%
\usepackage{hyperref}
\newcommand\id{\leavevmode\hbox{\small1\kern-3.3pt\normalsize1}}

\newcommand{\bra}{\langle}
\newcommand{\ket}{\rangle}

\newcommand{\tr}{\mbox{Tr}}

\begin{document}

\title{Bounding quantum correlations with indefinite causal order}


\author{{\v C}aslav Brukner$^{1,2}$}

\affiliation{
$^1$Faculty of Physics, University of Vienna, Boltzmanngasse 5,
A-1090 Vienna, Austria.\\
$^2$Institute for Quantum Optics and Quantum Information, Austrian Academy of Sciences, Boltzmanngasse 3,
A-1090 Vienna, Austria.}

\begin{abstract}
Causal inequalities are bounds on correlations obtained when operations take place in a causal sequence, i.e. in which the background time or definite causal structure pre-exists such that every operation is either in the future, in the past or space-like separated from any other operation. Recently, a framework was developed where quantum theory is assumed to be valid in local laboratories, but where no reference is made to any global causal relations between the operations in the laboratories. The framework was shown to allow for correlations that violate a bipartite causal inequality. Here we prove that the maximal violation of the causal inequality is upper bounded (analogously to the ``Tsirelson bound''`) under a restricted set of local operations involving binary observables. The bound is lower than what is algebraically possible.

\end{abstract}

\maketitle

The strength of violation of Bell's inequalities \cite{Bell}  in quantum theory is bounded from above by Tsirelson's bound~\cite{Cirelson}. The bound is a characteristic feature of quantum theory. One can conceive theories which violate Bell's inequalities stronger than quantum theory, still being in agreement with no signalling (i.e. not allowing instantaneous information transfer across space). A celebrated example are Popescu-Rohrlich boxes~\cite{PR}, also known as non-local boxes, which are hypothetical systems that achieve the algebraic bound for violation of the Bell inequality, yet do not allow signalling. Progress in understanding the difference between no-signalling classical, quantum, and superquantum correlations has provided insights into fundamental features of quantum theory, which demarcate it from both classical physics and more general probabilistic theories~\citep{dam,pawlowski,navascues,cabello,acin}.

In quantum mechanics and quantum field theory on curved space-times, all operations are embedded in a fixed space-time. Consequently, the correlations between operations respect definite causal order: they are either signalling correlations for the time-like or no-signalling correlations for the space-like separated operations. In quantum theory, however, every physical quantity is subject to quantum uncertainty. Recently, Hardy proposed to describe causal structures to be both dynamic, as in general relativity, as well as, indefinite, similarly to quantum observables~\cite{hardyqg}. This suggests that incorporating such causal structures into a quantum framework might lead to situations in which the causal ordering of events, as well as to whether their spatio-temporal distance is space-like or time-like is not fixed in advanced. The possibility of quantum computation on indefinite causal structures was proposed~\cite{hardyqgc} and it was shown that ``superposition of orders of quantum gates'' yield information processing advantages with respect to quantum circuits with a fixed order of gates~\cite{chiribella,chiribella1}, even in the asymptotic case~\cite{araujo}.

A framework was developed recently where it was presumed that operations in local laboratories are described by quantum theory (i.e. are completely-positive maps) while no reference is made to any global causal relations between these operations~\cite{OCB}. The surprising feature of this framework was the fact that there are situations where two operations are neither causally ordered nor in a probabilistic mixture of definite causal orders. This means that it is not possible to know the order between the two operations -- it is not clear whether one is prior to or after the other. Moreover, it was shown that the correlations between the operations violate a bipartite casual inequality, which is impossible if the operations are ordered according to a fixed causal structure. In the case of three parties the predefined causal order is violated by perfect signalling among the parties~\cite{amin} (this is analogue to the ``all versus nothing'' type of argument against local hidden variables~\cite{ghz}). Furthermore, classical correlations (i.e. locally compatible with classical probability theory) exist that allow for deterministic signalling between three or more parties incompatible with any predefined causal order~\cite{amin2}. For a bipartite case, however, violation of causal inequalities is possible only with quantum laboratories. Whether correlations violating causal inequalities are realizable in nature is an open question. See a review~\citep{caslav} on recent theoretical progress in the field. 

Here we derive a quantum bound (i.e. bound compatible with local quantum mechanics) on the strength of violation of the bipartite causal inequality. The proof is limited to the case where the parties perform operations from a restricted set of complete positive maps involving binary observables. Interestingly, both the causal and quantum bound on the causal inequality match numerically the corresponding classical and quantum (Tsirleson's) bound on the Clauser-Horne-Shimony-Holt version~\cite{CHSH} of Bell's inequality. 

In a general case, we have two scientists (who generally are dubbed Alice and Bob), who operate their experiment in two isolated laboratories. At a given run of the experiment the system enters into the respective laboratory of Alice and Bob respectively. Subsequently, Alice and Bob perform their operation on the system. And then, each experimenter transmits the system out of her/his laboratory. During the operations of each experimenter, the respective laboratories are completely shielded from the rest of the world. The system enters and exits each laboratory only {\it once} and is isolated from the outside world for the duration of the operations.

To describe the causal inequality, a game we might call ``guess your partner's input'' could be useful (see Figure 1, left)~\cite{OCB}. At the beginning each partner receives a random input bit value 0 or 1, say Alice receives the bit value $a$ and Bob $b$. In addition, Bob is given another random input $b'$, which specifies their goal: if $b'= 0$, Alice is required to return her best guess of Bob's input $b$, whereas if $b'=1$, he is asked to give his best guess for the bit $a$. We denote by $x$ and $y$ the guesses by Alice and Bob respectively, for the bit of the other. Their goal is to maximize the probability of success
\begin{gather}
p_{succ}:=\frac{1}{2}\left[ P(x=b|b'=0)+P(y=a|b'=1)\right]. \label{cave}
\end{gather}

It is easy to see that if all events obey causal order, the two partners cannot exceed the bound
\begin{gather}
p^{caus}_{succ}\leq 3/4.\label{bound}
\end{gather}
Without loss of generality consider the following case. The experiment performed by Alice is in the past of the one performed by Bob (see Figure 1 right). Accordingly, Alice is able to transmit her input $a$ to Bob. If $b'=1$, Alice and Bob have successfully concluded their experiment: in this case, Bob is able to expose Alice's bit: $y=a$. However, if she is asked to guess his bit, she cannot do it better than giving a random answer. This results in an overall success probability of 3/4. Formally, the inequality is based on a set of three assumptions: ``definite causal structure'', ``freedom of choice'' and ``closed laboratories'' (i.e. the laboratories are shielded from the rest of the world) which are thoroughly discussed in Ref.~\cite{OCB}. 

We now give a brief overview of the most important elements of the framework for quantum theory without a global causal order~\cite{OCB}, which are relevant for the present work. The main premiss of the framework is the validity of {\it local quantum mechanics: the local operations of each party are described by quantum theory,} i.e. the local operation are completely positive (CP) trace non-increasing maps. If an operation is performed on a quantum state $\rho$ and an outcome $i$ is observed, ${\cal M}_i(\rho)$ describes the updated state after the operation (up to normalization), where ${\cal M}_i$ is a CP trace non-increasing map. We denote Alice's map by ${\cal M}_i^A:{\cal L}({\cal H}^{A_1})\rightarrow {\cal L}({\cal H}^{A_2})$, where ${\cal L}({\cal H}^{A_1})$ is the space of matrices over the input Hilbert space ${\cal H}^{A_1}$ and ${\cal L}({\cal H}^{A_2})$ the one over the output Hilbert space ${\cal H}^{A_2}$. (The two Hilbert spaces can have different dimensions since the operation may involve ancillas). Similarly is for Bob's CP map. 

The Choi-Jamiolkowski (CJ) \cite{choi,jam} isomorphism allows to represent operations by operators (bipartite states) rather than maps. The CJ matrix $M^{A_1A_2}_i\in{\cal L}({\cal H}^{A_1}\otimes{\cal H}^{A_2})$ corresponding to a linear map ${\cal M}_i$ is defined as $M^{A_1A_2}_i:=\left[{\cal I}\otimes{\cal M}_i\left(|\phi^+\ket\bra \phi^+|\right)\right]^{\mathrm T}$, where $|\phi^+\ket=\sum_{j=1}^{d_{A_1}}|jj\ket \in {\cal H}^{A_1}\otimes{\cal H}^{A_2}$ is a (not normalized) maximally entangled state, the set of states $\left\{|j\ket\right\}_{j=1}^{d_{A_1}}$ is an orthonormal basis of ${\cal H}^{A_1}$ of dimension $d_{A_1}$, ${\cal I}$ is the identity map, and ${\mathrm T}$ denotes matrix transposition (the transposition is introduced for convenience).

The quantum framework with no reference to background causal structure leads to a certain form for the probability for a pair of Alice's and Bob's CP maps. It can be expressed as a bilinear function of the CJ representations $M^{A_1 A_2}_i$ and $M^{B_1 B_2}_j$ of local CP maps performed by Alice and Bob, respectively, as follows 
\begin{equation}
	\label{representation}
	P\left(\mathcal{M}^A_i, \mathcal{M}^B_j\right) = \tr \left[W^{A_1A_2B_1B_2}\left(M^{A_1A_2}_i\otimes M_j^{B_1B_2}\right)\right].
\end{equation}
Here $W^{A_1A_2B_1B_2} \in {\cal L}({\cal H}^{A_1}\otimes{\cal H}^{A_2}\otimes{\cal H}^{B_1}\otimes{\cal H}^{B_2})$ is a matrix from the space of matrices over the tensor product of the input and output Hilbert spaces. 

The $W$ matrix represents a new resource called ''process`` and is a generalization of the notion of ``state''. It can describe non-signalling correlations arising from measurements on a bipartite state, signalling ones, which can arise when a system is sent from one laboratory to another through a quantum channel, as well as correlations arising from the situations that are not causally ordered. The later type of \textit{qubit} process matrix (i.e. where all input and output Hilbert spaces are two-dimensional) was shown to allow the causal inequality to be violated by $(2+\sqrt{2})/4$, but the question has remained whether it is the highest possible value. Mathematically, a process matrix is a positive matrix and it returns unit probability for any pair of complete-positive trace-preserving maps. A closely related object to the process is a ``quantum comb''~\cite{networks}, but it is subject to additional conditions fixing a definite causal order.

\begin{figure}
\begin{center}
\includegraphics[width=8.8cm]{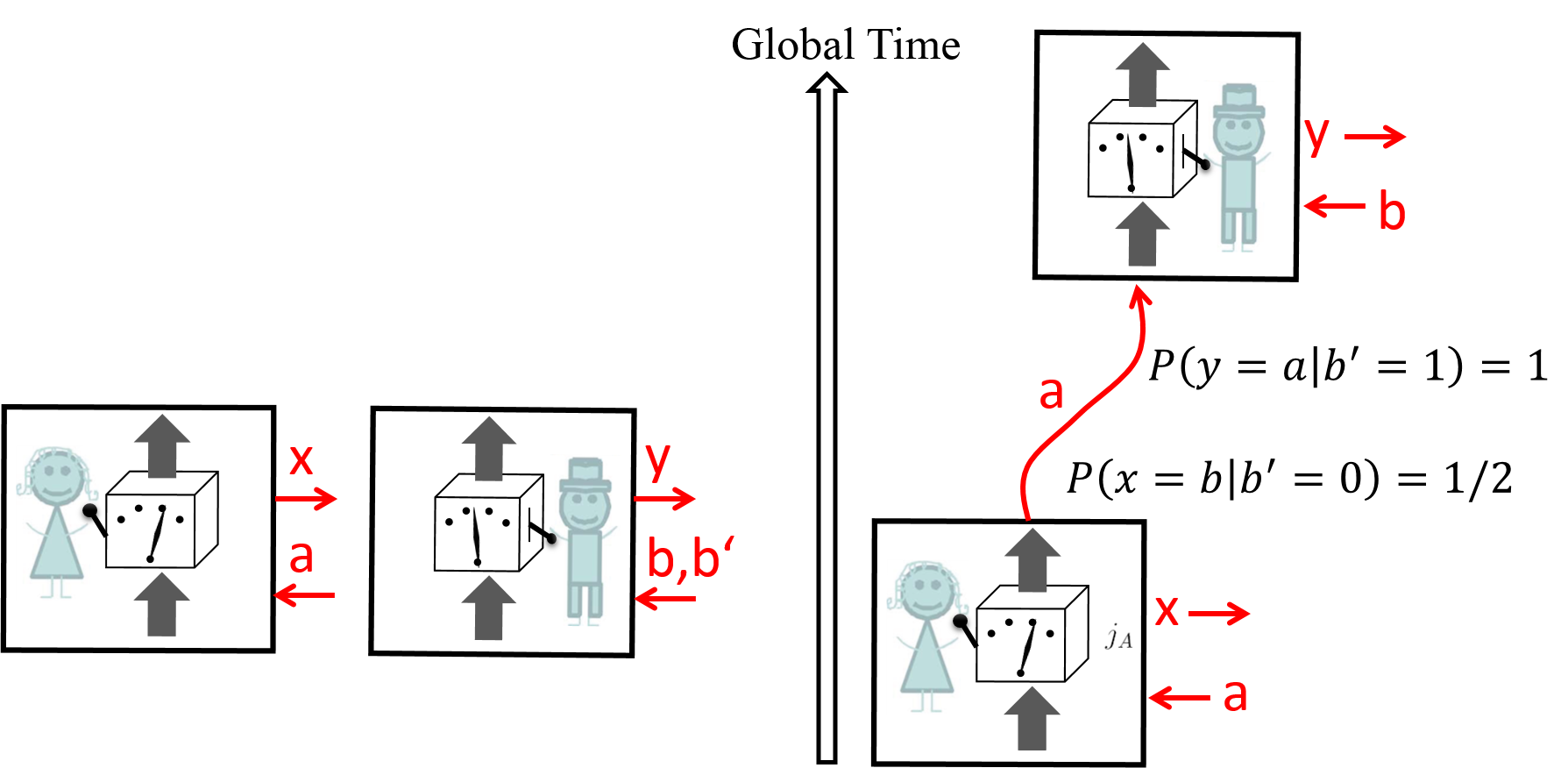}
\end{center}
\caption{(left) Causal game. Alice receives random bit $a$ and Bob receives random bits $b$ and $b'$. Depending on the bit value $b'$, either Alice will be asked to give her best guess of Bob's input $b$ ($b'=0$), or he will be required to guess her input $a$ ($b'=1$). (right) Strategy for accomplishing the game in a fixed causal structure. There exists a global background time according to which Alice's actions are strictly before Bob's. She sends her input $a$ to Bob, who can read it out and give his guess $y=a$. However, since her actions are in the past of Bob, she can only randomly guess his input bit $b$. This results in the probability of success $p_{succ}= \frac{1}{2}\left[ P(x=b|b'=0)+P(y=a|b'=1)\right]=3/4.$} \label{task}
\end{figure}

We now derive a quantum bound on violation of the causal inequality. We consider an arbitrary process matrix but restrict Alice's and Bob's CP maps to involve dichotomic observables with outcomes denoted by +1 and -1. Specifically, such a map could be of measurement-repreparation type, where the partners measure traceless dichotomic observables, and reprepare their states from a (degenerative) eigenspaces of the observables. A Hilbert-Schmidt basis of ${\cal L}({\cal H}^{X})$, with $X=A_1,A_2,B_1,B_2$ is given by a set of matrices $\{\sigma^X_{\mu}\}_{\mu=0}^{d_X^2-1}$, with $\sigma^X_0=\id_X$, $\tr \sigma^X_{\mu}\sigma^X_{\nu}=d_X\delta_{\mu\nu}$, and $\tr \sigma^X_{j}=0$ for $j=1,\dots ,d_X^2-1$ and $d_X$ is even. A traceless dichotomic observable $O_{\vec{m}}$ can be expressed as  $O_{\vec{m}}=\sum_{i>0} m_i \sigma^X_i \equiv (\vec{m}|\vec{\sigma})$, where $\vec{m}$ specifies the observable and $(...|...)$ is the scalar product in $d^2_X-2$ dimensional Euclidean space. Since $O_{\vec{m}}^2=\id$, one has $|\vec{m}|=1$, where $|...|$ denotes the two-norm. The projectors onto two eigenspaces are given as $P_{\vec{m}}^x = \frac{1}{2} [\id + (-1)^x (\vec{m}|\vec{\sigma})]$ with outcomes $x=0$ or $1$. We also introduce a traceless dichotomic ``correlation observable'' $O_{\hat{T}}=\sum_{i,j>0} T_{ij} \sigma^X_i \otimes \sigma^X_j \equiv (\hat{T}|\vec{\sigma}\otimes \vec{\sigma})$, where $O^2_{\hat{T}} = \id$ and $|\hat{T}|=1$.

We next consider a strategy for violation of the causal inequality. As Alice does not have an access to bit $b'$, she performs a single map decoding $x$ from the incoming system $A_1$ and encoding $a$ into $A_2$ and the correlations between systems $A_1$ and $A_2$. Hence, the possible operations performed by Alice can be represented by the CJ matrix as follows $\xi^{A_1A_2}(x,a)= \frac{1}{2d_{A_2}} \left[\id +(-1)^x (\vec{m}|\vec{\sigma}^{A_1})  + (-1)^F (\vec{n}|\vec{\sigma}^{A_2}) +(-1)^{F \oplus x} (\hat{T}|\vec{\sigma}^{A_1}\otimes \vec{\sigma}^{A_2})\right]$, where we omit the identity operators in the spaces of the subsystems. The encoding one-bit value function $F(x,a) \in \{0,1\}$ can depend both on $x$ and $a$. The sign $(-1)^{F\oplus x}$ is set to ensure the complete positivity of the CJ matrix in the particular case when the map is of the measurement-repreparation type: $ \xi^{A_1A_2}(x,a)=\frac{1}{2d_{A_2}} \left[\id+(-1)^x (\vec{m}|\vec{\sigma}) \right]^{A_1}\otimes\left[\id+ (-1)^{F} (\vec{n}|\vec{\sigma}) \right]^{A_2}$. In that case Alice measures observable $O_{\vec{m}}$ on the incoming system, assigning the value $x$ to the measured subspace $P_{\vec{m}}^x$. Subsequently, she reprepares the system, encoding $a$ in the totally mixed state $\frac{1}{d_{A_2}} P_{\vec{n}}^{F(x,a)}$ in the support of a subspace of another observable $O_{\vec{n}}$, and sends it out of her laboratory. In the most general case, $\vec{m}$, $\vec{n}$ and $\hat{T}$ can all depend on $a$ and $x$. The present maps belong to a restricted class. 

Bob chooses a strategy that depends on the bit value $b'$. If he wants to receive Alice's bit ($b'\!=\!1$), his encoding strategy is unimportant. Hence, the CJ matrix representing Bob's CP map can be chosen as $\eta_1^{B_1B_2}(y,b,b'\!=\!1) =  \frac{1}{2} \left[\id + (-1)^y (\vec{r}|\vec{\sigma})  \right]^{B_1} \! \otimes\rho^{B_2}$ (where for definiteness we denote the reprepared state by $\rho^{B_2}$). He  measures observable $O_{\vec{r}}$ on the incoming system and assigns $y$, to the outcome $P_{\vec{r}}^y$. If Bob wants to send his bit ($b'=0$), he applies the map with the CJ representation: 
$\eta^{B_1B_2}(y,b,b'=0)= \frac{1}{2d_{B_2}} \left[\id +(-1)^y (\vec{t}|\vec{\sigma}^{B_1})  + (-1)^G (\vec{o}|\vec{\sigma}^{B_2}) +(-1)^{G \oplus y} (\hat{S}|\vec{\sigma}^{B_1}\otimes \vec{\sigma}^{B_2})\right]$. Again this includes as a special case the maps of the measurement-repreparation type. Bob measures $O_{\vec{t}}$, assigning the value $y$ to the measured subspace $P_{\vec{t}}^y$.  He reprepares the state $\frac{1}{d_{B_2}} P_{\vec{o}}^{G(y,b)}$ from a subspace of the observable $O_{\vec{o}}$ of the outgoing system, where the encoding function $G(y,b)\in \{0,1\}$ can in general depend on both $b$ and the observed outcome $y$. Since in the case for $b'=0$ Bob sends a signal to Alice, his assignment in ${\cal H}^{B_1}$ is arbitrary. 

The most general bipartite process matrix has the form \cite{OCB}
\begin{align}
	W^{A_1A_2B_1B_2} 	&= \frac{1}{d_{A_1}d_{B_1}}\left(\id + \sigma^{B\preceq A}+ \sigma^{A\preceq B}+\sigma^{A\npreceq\nsucceq B}\right), \label{quantum} \\ \nonumber
	\sigma^{B\preceq A}	&:= \sum_{ij>0}c_{ij}\sigma^{A_1}_{i}\sigma^{B_2}_{j} + \sum_{ijk>0}d_{ijk}\sigma^{A_1}_{i}\sigma^{B_1}_{j}\sigma^{B_2}_{k},\\ \nonumber
	\sigma^{A\preceq B}	&:= \sum_{ij>0}e_{ij}\sigma^{A_2}_{i}\sigma^{B_1}_{j} + \sum_{ijk>0}f_{ijk}\sigma^{A_1}_{i}\sigma^{A_2}_{j}\sigma^{B_1}_{k},\\ \nonumber
	\sigma^{A\npreceq\nsucceq B} 	&:= \sum_{i>0}v_{i}\sigma^{A_1}_{i} + \sum_{i>0}x_{i}\sigma^{B_1}_{i} + \sum_{ij>0}g_{ij}\sigma^{A_1}_{i}\sigma^{B_1}_{j},\\ \nonumber
	\mbox{where }&	c_{ij}, d_{ijk}, e_{ij}, f_{ijk}, g_{ij}, v_{i}, x_{i} \in \mathbb {R}. 
\end{align}
The terms in $\sigma^{B\preceq A}$ allow signalling from Bob to Alice (channels from B to A), those in $\sigma^{A\preceq B}$ allow signalling from Alice to Bob (channels from A to B), whereas the terms $\sigma^{A\npreceq\nsucceq B}$ describe no-signalling situations (bipartite states). We will refer to terms of the form $\sigma^{A_1}_i \otimes \id^{rest}$ as of the type $A_1$, terms of the form $\sigma^{A_1}_i \otimes \sigma^{A_2}_j\otimes  \id^{rest}$ as of the type $A_1A_2$ and so on.

According to Eq.~\eqref{representation} the probabilities for different possible outcomes in the protocol are given by 
$P(xy|abb')= \tr [W^{A_1A_2B_1B_2}(\xi^{A_1A_2}(x,a)\eta^{B_1B_2}(y,b,b'))]$. For a successful calculation of the success probability, it is necessary to compute the intermediate probabilities $P(x|ab,b'\!=\!0) =$ $ \sum_y \tr [W^{A_1A_2B_1B_2}(\xi^{A_1A_2}(x,a)\eta^{B_1B_2}(y,b,b'\!=\!0))]$ as well as ${P(y|ab,b'\!=\!1) = \sum_x \tr [W^{A_1A_2B_1B_2} (\xi^{A_1A_2}(x,b)\eta^{B_1B_2}(y,b,b'\!=\!1))]}$. Note that the sum of a local CP map over the outcomes gives a complete-positive trace preserving map.

After a lengthy but straightforward calculation one obtains that either 
\begin{align}
	\label{aliceprob}	P(x\!=\!b|a,b,b'\!=\!0) &= \frac{1}{2} \left[1+ (-1)^b (\vec{v}|\vec{m}) + (\hat{c}|\vec{m} \otimes \vec{o})\right]  \mbox{ or} \\
	\label{aliceprob1} P(x\!=\!b|a,b,b'\!=\!0) &= \frac{1}{2} \left[1+ (-1)^b (\vec{v}|\vec{m}) + (\hat{d}|\vec{m} \otimes \hat{S}) \right].
\end{align}
Similarly one has either 
\begin{align}
	\label{bobprob}	P(y\!=\!a|a,b,b'\!=\!1) &= \frac{1}{2} \left[1+ (-1)^a (\vec{x}|\vec{r}) + (\hat{e}|\vec{n} \otimes \vec{r})\right]  \mbox{ or} \\
	\label{bobprob1} P(y\!=\!a|a,b,b'\!=\!1) &= \frac{1}{2} \left[1+ (-1)^a (\vec{x}|\vec{r}) + (\hat{f}|\hat{T} \otimes \vec{r} ) \right].
\end{align}
(One can obtain any sign $\pm$ in front of the third terms on the right-hand side, but it can always be absorbed by a choice of the vectors representing the local operations.) Here the tensors and vectors $\hat{c},\hat{d},\hat{e},\hat{f},\vec{v},\vec{x}$ are defined through their components $c_{ij},d_{ijk}, e_{ij}, f_{ijk}, v_{i}, x_{i}$ as introduced in~\eqref{quantum}, respectively. After averaging Eq.~\eqref{aliceprob} and \eqref{aliceprob1} over $b$, as well as Eq.~\eqref{bobprob} and \eqref{bobprob1} over $a$, and inserting different combinations of them in~\eqref{cave}, one obtains four possible expressions for the probability of success. In order to find the bound on the violation of the causal inequality we need to maximize the probability over the choice of observables. The four expressions for the maximal probability of success are given as follows: 
\begin{align}
p^{max}_{succ}&=\max_{\vec{n},\vec{r},\vec{m},\vec{o}} \frac{1}{4}  \left[ 2 + (\hat{e}|\vec{n} \otimes \vec{r}) +(\hat{c}|\vec{m} \otimes \vec{o})\right], \\
\label{nick}
p^{max}_{succ}&=\max_{\vec{n},\vec{r},\vec{m},\hat{S}} \frac{1}{4}  \left[ 2 + (\hat{e}|\vec{n} \otimes \vec{r}) +(\hat{d}|\vec{m} \otimes \hat{S})\right], \\
p^{max}_{succ}&=\max_{\hat{T},\vec{r},\vec{m},\vec{o}} \frac{1}{4}  \left[ 2 + (\hat{f}|\hat{T} \otimes \vec{r}) +(\hat{c}|\vec{m} \otimes \vec{o})\right], \\
p^{max}_{succ}&=\max_{\hat{T},\vec{r},\vec{m},\hat{S}} \frac{1}{4}  \left[ 2 + (\hat{f}|\hat{T} \otimes \vec{r}) +(\hat{d}|\vec{m} \otimes \hat{S})\right].
\end{align}

We give the proof for the case (\ref{nick}); other cases can be treated analogously. 
We first construct a properly normalized density matrix $\rho=\frac{1}{d_{A_2}d_{B_2}} W^{A_1A_2B_1B_2}$ from the given process matrix. Next we introduce vectors $\vec{a}\equiv (\id \otimes O_{\vec{n}} \otimes O_{\vec{r}} \otimes \id) \sqrt{\rho}$, $\vec{b}\equiv (O_{\vec{m}} \otimes \id \otimes O_{\hat{S}}) \sqrt{\rho}$ and $\vec{i}\equiv (\id \otimes \id \otimes \id \otimes \id) \sqrt{\rho}$ from the complex vector space of dimension $(d_{A_1} d_{A_2}d_{B_1}d_{B_2})^2$, where we consider the operators on the right hand sides as vectors. The scalar product between two vectors $\vec{c}\equiv C$ and $\vec{d} \equiv D$ is defined by the inner product of the corresponding matrices:  $(\vec{c}|\vec{d}) \equiv \mbox{Tr}\left(C^{\dagger}D\right)$. The vectors are of the unit length since $(\vec{a}|\vec{a})= \tr \left[(\id \otimes O_{\vec{n}} \otimes O_{\vec{r}} \otimes \id) \sqrt{\rho} \sqrt{\rho}(\id \otimes O_{\vec{n}} \otimes O_{\vec{r}} \otimes \id) \right] =\tr (\rho)=1$. Similarly, one has $(\vec{b}|\vec{b})=(\vec{i}|\vec{i})=1$. On the other hand, $(\vec{a}|\vec{i})=\tr \left[ (\id \otimes O_{\vec{n}} \otimes O_{\vec{r}} \otimes \id) \rho \right] =(\hat{e}|\vec{n} \otimes \vec{r})$ and $(\vec{b}|\vec{i}) = \tr \left[(O_{\vec{m}} \otimes \id \otimes O_{\hat{S}}) \rho \right] = (\hat{d}|\vec{m} \otimes \hat{S} )$. Finally, $(\vec{a}|\vec{b})=\tr\left[O_{\vec{m}} \otimes O_{\vec{n}} \otimes O_{\hat{S}}(O_{\vec{r}}\otimes \id) \rho \right] =0$ since $\rho$ contains neither terms of the type $A_1A_2B_2$ nor $A_1A_2B_1B_2$.

We can now rewrite Eq.~\eqref{nick} in terms of the vectors $\vec{a}, \vec{b}$ and $\vec{i}$ as follows: 
\begin{gather}
p^{max}_{succ}=\max_{\vec{a},\vec{b}} \frac{1}{4}  \left[ 2 + (\vec{a}|\vec{i}) + (\vec{b}|\vec{i}) \right].
\end{gather}
Note that $\vec{a}$ and $\vec{b}$ cannot be chosen independently of $\vec{i}$ as they all depend on the given $\rho$. Since the three vectors are of unit length and $\vec{a}$ and $\vec{b}$ are orthogonal to each other, the maximum is achieved if $\vec{i}$ lies in the plane spanned by $\vec{a}$ and $\vec{b}$. Then one can write $(\vec{a}|\vec{i})=\cos \theta$ and $(\vec{b}|\vec{i})=\sin \theta$. This finally implies
\begin{gather}
p^{max}_{succ} \leq \max_{\theta} \frac{1}{4}  (2 + \cos \theta + \sin \theta) = \frac{1}{2} (1 + \frac{1}{\sqrt{2}}).
\label{qbound}
\end{gather}

As shown in Ref.~\cite{OCB} the bound can be achieved with the qubit process matrix 
\begin{equation}
\label{quantum1}
	\!\!\!	W^{A_1A_2B_1B_2}=\frac{1}{4}\left[\id^{A_1A_2B_1B_2} + \frac{1}{\sqrt{2}}\left(\sigma_z^{A_2}\sigma_z^{B_1} + \sigma_z^{A_1}\sigma_x^{B_1}\sigma_z^{B_2} \right) \right],
\end{equation}
where $A_1$, $A_2$, $B_1$, and $B_2$ are two-level systems (e.g. the spin degrees of freedom of a spin-$\frac{1}{2}$ particle) and $\sigma_x$ and $\sigma_z$ are the Pauli spin matrices. The strategies are of the measurement-repreparation type. Alice always performs map $\frac{1}{4} [\id +(-1)^x \sigma_z]^{A_1}\otimes [\id + (-1)^a \sigma_z]^{A_2}$. If $b'=1$, Bob chooses map $\frac{1}{2} [\id +(-1)^y \sigma_z]^{B_1}\otimes \rho^{B_2}$, while if $b'=0$, he performs $\frac{1}{4} [\id +(-1)^y \sigma_x]^{B_1}\otimes [\id + (-1)^{y\oplus b} \sigma_z]^{B_2}$.

Signalling correlations do not have a definite causal order if they violate a causal inequality. Another feature of quantum theory without global causal order is the existence of \textit{causally non-separable} $W$ processes, i.e. processes which cannot be written in a causal form (or, more generally, as a mixture of processes in a causal form), $W \neq \lambda W^{A\npreceq B} + (1-\lambda) W^{B \npreceq A}$, where $0 \leq \lambda \leq 1$, and $W^{A\npreceq B}$ is the process in which Alice cannot signal to Bob (i.e. channels from Bob to Alice or the two share a bipartite state) and $W^{B \npreceq A}$ in which Bob cannot signal to Alice~\cite{OCB}. If the process matrix can be written in a causal form, we call it causally separable. The process matrix~\eqref{quantum1} is an example of a causally non-separable process. 

The two notions are analogous to nonlocality and entanglement in quantum theory. Even though the latter are intimately related, they are distinct notions. There are mixed quantum states (for example, the Werner states~\cite{werner}), which, while entangled, yield outcomes that allow a description in terms of local hidden variables. Next we consider a family of processes (which are process analogue to the Werner states) and show that they violate the causal inequality if and only if they are causally non-separable. Therefore, there is no gap between causal non-separability and correlations violating the causal inequality for this family of processes.

We consider the processes of the form: 
\begin{equation}
W_{\eta_1,\eta_2}=\frac{1}{4}\left[\id^{A_1A_2B_1B_2} + \eta_1 \sigma_z^{A_2}\sigma_z^{B_1} + \eta_2 \sigma_z^{A_1}\sigma_x^{B_1}\sigma_z^{B_2}\right].
\label{werner}
\end{equation}
Since the process matrix is positive, one has $\eta^2_1 +\eta^2_2 \leq 1$. For the special case of $\eta_1=\eta_2 \equiv \eta \geq 0$, the process can be written as an admixture of totally noisy process $W_{noise}= \frac{1}{4}\id^{A_1A_2B_1B_2}$ to the process $W$ from Eq.~\eqref{quantum1}: $W_{\xi}=\xi W + (1-\xi) W_{noise}$ with $\xi=\eta \sqrt{2}$.

The process matrix is causally non-separable if and only if its distance to the closest causally separable process is strictly positive. As for the distance measure we introduce the following geometric measure
\begin{equation}
D(W) \equiv \mbox{min}_{W_c \in \Omega_c} ||W - W_c||^2,
\end{equation}
where $\Omega_c$ is the set of causally separable processes and $|| W - W_c||^2 = \mbox{Tr}[(W-W_c)^2]$ is the square norm in the Hilbert-Schmidt space.  

We decompose any causally separable process matrix as $W_c = \lambda W^{A \preceq B} + (1-\lambda) W^{A \npreceq B}$, where $W^{A \preceq B}$ contains only terms $\sigma^{A\preceq B}$ from Eq.~\eqref{quantum1}, whereas $W^{A \npreceq B}$ contains both terms $\sigma^{A\preceq B}$ and $\sigma^{A\npreceq\nsucceq B}$. This particular decomposition is chosen because in the Hilbert-Schmidt space $W^{A \preceq B}$ and $ W^{A \npreceq B}$ are  orthogonal to each other on the traceless parts. To determine the geometric distance we compute the following: $\mbox{Tr}W^2=1 + \eta^2_1 + \eta^2_2$, $\mbox{Tr}W^2_c= \lambda^2 (1 + e^2_{zz} + \Gamma^2_1) + 2 \lambda (1-\lambda) + (1-\lambda)^2 (1 + d^2_{zxz} + \Gamma^2_2)$ and $\mbox{Tr} (W W_c) = \lambda (1+\eta_1 e_{zz}) + (1-\lambda) (1+\eta_2 d_{zxz})$. Here $\Gamma^2_1$ denotes the sum of the squares of all but $e^2_{zz}$ element in the Hilbert-Schmidt decomposition of $W^{A \preceq B}$. Similarly, $\Gamma^2_2$ is the sum of the squares of all but $d^2_{zxz}$ element in the decomposition of $W^{B \npreceq A}$.

\begin{figure}
\begin{center}
\includegraphics[width=6cm]{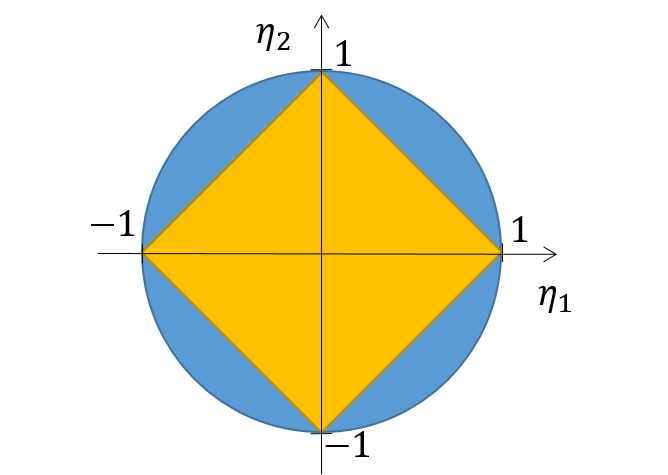}
\end{center}
\caption{Parameter space of $\eta_1$ and $\eta_2$ of process (\ref{werner}). The positive process matrices correspond to the entire of the circle $\eta_1^2 + \eta_1^2 \leq 1$. The blue region represents causally non-separable processes; the yellow one $|\eta_1| + |\eta_2| \leq 1$ causally separable ones.} \label{parameter}
\end{figure}

The geometric distance to the set of causally separable processes can be expressed as 
\begin{equation}
D(W)= \mbox{min}_{\vec{x}} || \vec{x} - \vec{y}||^2,
\end{equation}
where $\vec{x} \equiv (\lambda e_{zz}, (1-\lambda) d_{zxz})$ and  $\vec{y} \equiv (\eta_1,\eta_2)$ and $\Gamma_1=\Gamma_2=0$ is taken because otherwise $D(W)>0$ always. Hence, the geometric distance will be zero if there is $\vec{x}=\vec{y}$, i.e. if $\eta_1=\lambda e_{zz}$ and $\eta_2=(1-\lambda) d_{zxz}$. Note that due to $\Gamma_1=\Gamma_2=0$, one has $|e_{zz}|,|d_{zxz}| \leq 1$ to guarantee the positivity of $W^{A \preceq B}$ and $W^{B \npreceq A}$, respectively. This implies that whenever the distance is zero one has $1 = \lambda + (1-\lambda) = \frac{\eta_1}{e_{zz}}+\frac{\eta_2}{d_{zxz}} \geq |\eta_1| + |\eta_2|$, where the sign of $e_{zz}$ and $d_{zxz}$ is chosen to match those of $\eta_1$ and $\eta_2$, respectively. Hence, if $ |\eta_1| + |\eta_2| > 1 $, the distance is always strictly positive, and the process is causally non-separable. The opposite is also true, since at the values $ |\eta_1| + |\eta_2| = 1 $ one can always write the process matrix in a causally separable form as $W_{\eta_1,\eta_2}= |\eta_1| W^{A \preceq B} + |\eta_2| W^{B \preceq A}$ with $W^{A \preceq B}=\frac{1}{4} [\id^{A_1A_2B_1B_2} + \mbox{sgn}[\eta_1] \sigma_z^{A_2}\sigma_z^{B_1}]$ and
$W^{B \preceq A}= \frac{1}{4} [\id^{A_1A_2B_1B_2} +\mbox{sgn}[\eta_2] \sigma_z^{A_1}\sigma_x^{B_1}\sigma_z^{B_2}]$, where $\mbox{sgn}[...]$ is the sign of $[...]$. On the other hand, the causal inequality is violated if $|\eta_1| + |\eta_2| >1$, as one can easily convince oneself.

In conclusion, under restricted class of local strategies involving binary observables the quantum bound on the violation of the causal inequality is $\frac{1}{2} (1 + \frac{1}{\sqrt{2}})$. An intuitive understanding of this violation becomes possible through the realization that Bob, with a probability of $\sqrt{2}/2$ respectively, might end up either ``before'' or ``after'' Alice. He, however, cannot choose the causal order with certainty, as the bound is lower than what is algebraically maximal possible value. This raises the question of why it is so, and what principle limits quantum correlations with indefinite causal order.

\acknowledgements{I thank M. Araujo, F. Costa, A. Feix and I. Ibnouhsein for discussions. This work has been supported by the European Commission Project RAQUEL, the John Templeton Foundation, FQXi, and the Austrian Science Fund (FWF) through CoQuS, SFB FoQuS, and the Individual Project 2462.}

\end{document}